\documentstyle[12pt]{article}

\catcode`\@=11
\def\@citex[#1]#2{%
\if@filesw \immediate \write \@auxout {\string \citation {#2}}\fi
\@tempcntb\m@ne \let\@h@ld\relax \def\@citea{}%
\@cite{%
  \@for \@citeb:=#2\do {%
    \@ifundefined {b@\@citeb}%
      {\@h@ld\@citea\@tempcntb\m@ne{\bf ?}%
      \@warning {Citation `\@citeb ' on page \thepage \space undefined}}%
      {\@tempcnta\@tempcntb \advance\@tempcnta\@ne%
      \@tempcntb\number\csname b@\@citeb \endcsname \relax%
      \ifnum\@tempcnta=\@tempcntb 
	\ifx\@h@ld\relax%
	  \edef \@h@ld{\@citea\csname b@\@citeb\endcsname}%
	\else%
	  \edef\@h@ld{\ifmmode{-}\else--\fi\csname b@\@citeb\endcsname}%
	\fi%
      \else
	\@h@ld\@citea\csname b@\@citeb \endcsname%
	\let\@h@ld\relax%
      \fi}%
    \def\@citea{,\penalty\@highpenalty\,}%
  }\@h@ld
}{#1}}

\def\@citeb#1#2{{[#1]\if@tempswa , #2\fi}}
\def\@citeu#1#2{{$^{#1}$\if@tempswa , #2\fi }}
\def\@citep#1#2{{#1\if@tempswa , #2\fi}}

\def\bcites{         
	\catcode`\@=11
	\let\@cite=\@citeb
	\catcode`\@=12
}

\def\upcites{         
	\catcode`\@=11
	\let\@cite=\@citeu
	\catcode`\@=12
}

\def\plaincites{      
	\catcode`\@=11
	\let\@cite=\@citep
	\catcode`\@=12
}

\newcount\hour
\newcount\minute
\newtoks\amorpm
\hour=\time\divide\hour by 60
\minute=\time{\multiply\hour by 60 \global\advance\minute by-\hour}
\edef\standardtime{{\ifnum\hour<12 \global\amorpm={am}%
	\else\global\amorpm={pm}\advance\hour by-12 \fi
	\ifnum\hour=0 \hour=12 \fi
	\number\hour:\ifnum\minute<10 0\fi\number\minute\the\amorpm}}
\edef\militarytime{\number\hour:\ifnum\minute<10 0\fi\number\minute}

\def\draftlabel#1{{\@bsphack\if@filesw {\let\thepage\relax
   \xdef\@gtempa{\write\@auxout{\string
      \newlabel{#1}{{\@currentlabel}{\thepage}}}}}\@gtempa
   \if@nobreak \ifvmode\nobreak\fi\fi\fi\@esphack}
	\gdef\@eqnlabel{#1}}
\def\@eqnlabel{}
\def\@vacuum{}
\def\marginnote#1{}
\def\draftmarginnote#1{\marginpar{\raggedright\scriptsize\tt#1}}
\def\draft{
	\pagestyle{plain}
	\overfullrule=2pt
	\oddsidemargin -.5truein
	\def\@oddhead{\sl \phantom{\today\quad\militarytime} \hfil
	\smash{\Large\sl DRAFT} \hfil \today\quad\militarytime}
	\let\@evenhead\@oddhead
	\let\label=\draftlabel
	\let\marginnote=\draftmarginnote
	\def\ps@empty{\let\@mkboth\@gobbletwo
	\def\@oddfoot{\hfil \smash{\Large\sl DRAFT} \hfil}
	\let\@evenfoot\@oddhead}
	\def\@eqnnum{(\theequation)\rlap{\kern\marginparsep\tt\@eqnlabel}%
	\global\let\@eqnlabel\@vacuum}  }

\def\blackfonts{
	\font\blackboard=msbm10 scaled\magstep1
	\font\blackboards=msbm8
	\font\blackboardss=msbm6
}

\def\yblack{
	\blackfonts
	\newfam\black
	\textfont\black=\blackboard
	\scriptfont\black=\blackboards
	\scriptscriptfont\black=\blackboardss
	\def\ZZ{{\fam\black\relax Z}}
	\def\NN{{\fam\black\relax  N}}
	\def\CC{{\fam\black\relax  C}}
	\def\RR{{\fam\black\relax  R}}
	\def\SS{{\fam\black\relax  S}}
	\def\QQ{{\fam\black\relax  Q}}
	\def\PP{{\fam\black\relax  P}}
}

\def\nblack{            
    \def\ZZ{{Z \mskip-10mu Z}}
	\def\NN{{N \mskip-14mu N}}
	\def\CC{{C \mskip-11mu C}}
	\def\RR{{R \mskip-11mu R}}
	\def\SS{{\bf S}}
	\def\QQ{{Q \mskip-12mu Q}}
	\def\PP{{P \mskip-11mu P}}
}

\def\prep{         
	\catcode`\@=11
	\input art10.sty
	\catcode`\@=12
	
	\let\small\null
	\def\blackfonts{
		\font\blackboard=msbm10
		\font\blackboards=msbm7
		\font\blackboardss=msbm5
	}
	\let\sl\it
	\twocolumn
	\sloppy
	\voffset=-2.54truecm
	\hoffset=-2.54truecm
	\flushbottom
	\parindent 1em
	\leftmargini 2em
	\leftmarginv .5em
	\leftmarginvi .5em
	\marginparwidth 48pt
	\marginparsep 10pt
	\setlength{\columnsep}{2truecm}
	\setlength{\textwidth}{25.4truecm}
	\setlength{\textheight}{17truecm}
	\baselineskip=16pt
	\oddsidemargin .18truein
	\evensidemargin .17truein
}

\def\eqalign#1{\null\,\vcenter{\openup\jot\m@th
  \ialign{\strut\hfil$\displaystyle{##}$&$\displaystyle{{}##}$\hfil
      \crcr#1\crcr}}\,}
\def\eqalignno#1{\displ@y \tabskip\centering
  \halign to\displaywidth{\hfil$\@lign\displaystyle{##}$\tabskip\z@skip
    &$\@lign\displaystyle{{}##}$\hfil\tabskip\centering
    &\llap{$\@lign##$}\tabskip\z@skip\crcr
    #1\crcr}}

\def\section{\@startsection {section}{1}{\z@}{3.ex plus 1ex minus
 .2ex}{2.ex plus .2ex}{\large\bf}}
\def\subsection{\@startsection{subsection}{2}{\z@}{2.75ex plus 1ex minus
 .2ex}{1.5ex plus .2ex}{\bf}}

\def\appendix{{\newpage\section*{Appendix}}\let\appendix\section%
	{\setcounter{section}{0}
	\gdef\thesection{\Alph{section}}}\section}

\def\abstract{\if@twocolumn
\section*{Abstract}
\else 
\begin{center}
{\bf Abstract\vspace{-.5em}\vspace{0pt}}
\end{center}
\quotation
\fi}

\catcode`\@=12

\def\noj#1,#2,{{\bf #1} (19#2)\ }
\def\jou#1,#2,#3,{{\sl #1\/ }{\bf #2} (19#3)\ }
\def\ann#1,#2,{{\sl Ann.\ Physics\/ }{\bf #1} (19#2)\ }
\def\cmp#1,#2,{{\sl Comm.\ Math.\ Phys.\/ }{\bf #1} (19#2)\ }
\def\invm#1,#2,{{\sl Invent.\ Math.\/ }{\bf #1} (19#2)\ }
\def\actm#1,#2,{{\sl Acta.\ Math.\/ }{\bf #1} (19#2)\ }
\def\jdg#1,#2,{{\sl J.\ Diff.\ Geom.\/ }{\bf #1} (19#2)\ }
\def\cq#1,#2,{{\sl Class.\ Quantum Grav.\/ }{\bf #1} (19#2)\ }
\def\sm#1,#2,{{\sl Selec.\ Math.\/ }{\bf #1} (19#2)\ }
\def\ijmp#1,#2,{{\sl Int.\ J.\ Mod.\ Phys.\/ }{\bf A#1} (19#2)\ }
\def\jmp#1,#2,{{\sl J.\ Math.\ Phys.\/ }{\bf #1} (19#2)\ }
\def\jmphy#1,#2,{{\sl J.\ Geom.\ Phys.\/ }{\bf #1} (19#2)\ }
\def\grg#1,#2,{{\sl Gen.\ Rel.\ Grav.\/ }{\bf #1} (19#2)\ }
\def\mpl#1,#2,{{\sl Mod.\ Phys.\ Lett.\/ }{\bf A#1} (19#2)\ }
\def\nc#1,#2,{{\sl Nuovo Cim.\/ }{\bf #1} (19#2)\ }
\def\np#1,#2,{{\sl Nucl.\ Phys.\/ }{\bf B#1} (19#2)\ }
\def\pl#1,#2,{{\sl Phys.\ Lett.\/ }{\bf #1B} (19#2)\ }
\def\pla#1,#2,{{\sl Phys.\ Lett.\/ }{\bf #1A} (19#2)\ }
\def\pr#1,#2,{{\sl Phys.\ Rev.\/ }{\bf #1} (19#2)\ }
\def\prd#1,#2,{{\sl Phys.\ Rev.\/ }{\bf D#1} (19#2)\ }
\def\prl#1,#2,{{\sl Phys.\ Rev.\ Lett.\/ }{\bf #1} (19#2)\ }
\def\prp#1,#2,{{\sl Phys.\ Rept.\/ }{\bf #1C} (19#2)\ }
\def\ptp#1,#2,{{\sl Prog.\ Theor.\ Phys.\/ }{\bf #1} (19#2)\ }
\def\ptpsup#1,#2,{{\sl Prog.\ Theor.\ Phys.\/ Suppl.\/ }{\bf #1} (19#2)\ }
\def\rmp#1,#2,{{\sl Rev.\ Mod.\ Phys.\/ }{\bf #1} (19#2)\ }
\def\yadfiz#1,#2,#3[#4,#5]{{\sl Yad.\ Fiz.\/ }{\bf #1} (19#2) #3%
\ [{\sl Sov.\ J.\ Nucl.\ Phys.\/ }{\bf #4} (19#2) #5]}
\def\zh#1,#2,#3[#4,#5]{{\sl Zh.\ Exp.\ Theor.\ Fiz.\/ }{\bf #1} (19#2) #3%
\ [{\sl Sov.\ Phys.\ JETP\/ }{\bf #4} (19#2) #5]}

\def\beq{\begin{equation}}
\def\eeq{\end{equation}}
\def\beqar{\begin{eqnarray}}
\def\eeqar{\end{eqnarray}}

\def\nfrac#1#2{{\displaystyle{\vphantom1\smash{\lower.5ex\hbox{\small$#1$}}%
	\over\vphantom1\smash{\raise.25ex\hbox{\small$#2$}}}}}

\def\p#1{\mskip#1mu}
\def\n#1{\mskip-#1mu}
\def\stop{\p6.}
\def\comma{\p6,}

\def\to{\rightarrow}

\def\lae{\mathrel{\mathop{\smash{\lower .5 ex \hbox{$\stackrel<\sim$}}}}}
\def\lae{\mathrel{\mathop{\smash{\lower .5 ex \hbox{$\stackrel>\sim$}}}}}

\def\pa{\partial}

\def\l:{\mathopen{:}\,}
\def\r:{\,\mathclose{:}}

\def\[{\left[}          \def\]{\right]}
\def\({\left(}          \def\){\right)}
\def\<{\left<}          \def\>{\right>}

\def\a{\alpha}
\def\b{\beta}

\def\m{\mu}
\def\n{\nu}
\def\r{\rho}
\def\s{\sigma}
\def\g{\gamma}
\def\G{\Gamma}

\catcode`\@=11
\def\theequation{\arabic{equation}}
\catcode`\@=12

\nblack
\bcites

\nblack

\catcode`\@=11
\def\theequation{\thesection.\arabic{equation}}
\@addtoreset{equation}{section}
\@addtoreset{footnote}{section}
\@addtoreset{footnote}{subsection}
\catcode`\@=12

\yblack

\newcommand{\beqn}{\begin{equation}}
\newcommand{\eeqn}{\end{equation}}
\newcommand{\beqnarray}{\begin{eqnarray}}
\newcommand{\eeqnarray}{\end{eqnarray}}
\begin{document}
\begin{titlepage}

\begin{center}
August 18, 1996
\hfill DUKE-TH-96-124 \\
\hfill LBNL-39156, UCB-PTH-96/33 \\
\hfill NSF-ITP-96-65 \\
\hfill                  WIS-96/34/July-PH               \\
\hfill                  hep-th/9608116

\vskip 1 cm
{\large \bf Supersymmetric Cycles in Exceptional Holonomy Manifolds and
Calabi-Yau 4-Folds\\}
\vskip 1 cm
{Katrin Becker$^1$, Melanie Becker$^2$,
David R. Morrison$^3$, Hirosi Ooguri$^{4,5}$,\\
 Yaron Oz$^6$,
and Zheng Yin$^{4,5}$ }\\
\vskip 0.7cm

{\small
$^1$ {\it Institute for Theoretical Physics, University of California\\
Santa Barbara, CA 93106-4030, USA\\}
\smallskip
$^2$ {\it Department of Physics, University of California,
Santa Barbara, CA 93106-9530, USA\\}
\smallskip
$^3$ {\it Department of Mathematics,
Duke University, Box 90320, Durham, NC 27708, USA\\}
\smallskip
$^4$ {\it Department of Physics,
University of California, Berkeley, CA 94720, USA\\}
\smallskip
$^5$ {\it Theoretical Physics Group,
Ernest Orlando Lawrence Berkeley National Laboratory\\
University of California, Berkeley, CA 94720, USA\\}
\smallskip
$^6$ {\it Department of Particle Physics, Weizmann Institute of Science
 76100 Rehovot Israel}}

\vskip 0.5 cm
\begin{abstract}
We derive in the SCFT and low energy effective action frameworks
the  necessary and sufficient conditions for
supersymmetric cycles in exceptional holonomy manifolds and
Calabi-Yau 4-folds.
We show that the Cayley cycles in $Spin(7)$ holonomy eight-manifolds
and the associative and coassociative
cycles in
$G_2$ holonomy
seven-manifolds preserve half of the space-time supersymmetry.
We find that
while the holomorphic and special Lagrangian cycles in Calabi-Yau
4-folds preserve half of the space-time supersymmetry,
the Cayley submanifolds  are novel
as they preserve only one quarter of it.
We present some simple examples. Finally, we discuss the implications
of these supersymmetric cycles on mirror symmetry in higher dimensions.

\end{abstract}

\end{center}
\end{titlepage}

\section{Introduction}

A supersymmetric cycle is characterized by the property that the worldvolume
theory of a brane wrapping around it is supersymmetric.
The conditions for supersymmetric cycles in Calabi-Yau 3-folds
have been analyzed using the low
energy effective actions for branes \cite{BBS,bsv}, where two types of
conditions have been found. The first type
corresponds to even-dimensional cycles being complex (holomorphic)
submanifolds, i.e.,
having $\frac{1}{p!}k^p$ as their volume form, where $k$ denotes the
K\"ahler  form. The second type corresponds to middle-dimensional cycles
being special Lagrangian, i.e., Lagrangian submanifolds having
$Re({\Omega})$ as their volume form
where $\Omega$ corresponds to the nowhere vanishing holomorphic $(n,0)$
form on the Calabi-Yau $n$-fold.
The special Lagrangian and complex cycles in Calabi-Yau 3-folds and
4-folds have been
shown in \cite{OOY}
to arise from the large volume limit of $N=2$ SCFT boundary conditions of
$A$ and $B$ types respectively. Both types break half of the space-time
supersymmetry.

Our aim is to study supersymmetric cycles of exceptional type that are not
complex or special Lagrangian submanifolds, which
exist in $Spin(7)$, $SU(4)$ and $G_2$ holonomy manifolds.
For that it will be useful to introduce the concept of calibration
\cite{HL} which is the appropriate framework to study supersymmetric cycles.
A calibration is a closed $p$-form $\varphi$
on a Riemannian manifold of dimension $n$,
such that its restriction to each tangent $p$-plane of $M$ is less or equal
to the volume of the plane. Submanifolds for which there is equality are
said to be calibrated by $\varphi$.
A calibrated submanifold has the least volume in its homology class.
In fact, this property of a calibration provides a natural geometric
interpretation of the Bogomolnyi bound for D-branes wrapped about such
submanifolds, with the calibrated submanifolds corresponding to the BPS states
which saturate the bound.
Complex and special Lagrangian submanifolds are calibrated by
$\frac{1}{p!}k^p$ and $Re({\Omega})$ respectively.
In addition to these calibrations there exist exceptional ones \cite{HL}.
The Cayley calibration is a self-dual 4-form
on eight-dimensional manifolds
with holonomy contained in $Spin(7)$. The associative calibration is a 3-form
on seven dimensional manifolds with holonomy contained in $G_2$, and the
coassociative calibration is its Hodge dual.

In this paper we will analyze the supersymmetric cycles associated with
these exceptional calibrations using the SCFT framework and the low energy
effective action approach.
In section 2 we will consider the Cayley calibration
in $Spin(7)$. In section 3 we consider
$SU(4)$ holonomy eight-manifolds and the
associative and
coassociative calibrations in $G_2$ holonomy seven-dimensional manifolds
are discussed in section 4.
We will construct the SCFT boundary conditions
which in the large volume limit are associated with these cycles. 
We will find that
the Cayley 4-cycle in $SU(4)$ holonomy Calabi-Yau 4-fold is novel as
it preserves only one quarter of space-time supersymmetry, 
while all the others
preserve as usual half of the supersymmetry.
Using the supersymmetry transformations
of the low energy effective action for branes compactified on the Calabi-Yau
4-fold we derive the necessary and sufficient conditions for
supersymmetric cycles. As expected,
these conditions will coincide with the large volume
limit of the SCFT boundary conditions.
We present some simple examples of supersymmetric cycles in Calabi-Yau
4-folds.
In section 5 we discuss the implications of these supersymmetric
cycles on mirror
symmetry in higher dimensions.

\section{$Spin(7)$ holonomy}

Let $M$ be an eight-manifold. A $Spin(7)$ structure on $M$ is given by
a closed self-dual $Spin(7)$ invariant
4-form $\Phi$. This defines a metric $g$ with
holonomy group $Hol(g) \subset Spin(7)$. Such a metric is Ricci-flat.
Compact $Spin(7)$ holonomy manifolds have been constructed in \cite{Joyce3}
by resolving the singularities of $T^8/\Gamma$ orbifolds. Here $T^8$ is
equipped with a flat $Spin(7)$ structure and $\Gamma$ is a finite group
of  isometries of $T^8$  preserving that structure.
On a $Spin(7)$ holonomy manifold there exists one covariantly constant
spinor, which will provide us, upon compactification, with one space-time
supersymmetry.

The 4-form $\Phi$ can be used as a calibration called the Cayley calibration.
The calibration in general
is related to the covariantly constant spinor
via squaring \cite{H} which basically means that the calibration form can
be constructed from an appropriate product of two spinors.


The extended symmetry algebra of sigma models on $Spin(7)$ manifolds
has been found in \cite{sv}.
In addition to the stress momentum tensor $T$ and its superpartner $G$,
it contains two operators $\tilde{X}$ and $\tilde{M}$ with spins 2 and
$\frac{3}{2}$ respectively.
The presence of the spin 2 operator $\tilde{X}$ may be understood along the
following lines:  Recall that corresponding to the covariantly constant spinor
there exists a dimension $\frac{1}{2}$ Majorana-Weyl
spectral flow
operator $\Psi_L$
mapping the Neveu-Schwarz (NS) sector to the Ramond sector.
It implies
the existence of a dimension 2 operator
 $\tilde{X}$,
which is the
energy-momentum tensor for the $c=\frac{1}{2}$ Majorana-Weyl fermion
(Ising model),
mapping the NS to NS sectors.
In the large volume limit of the manifold $M$,  $\tilde{X}$ takes the
form \cite{sv}
\beq
\tilde{X}_L = \frac{1}{2}g_{\mu\nu} \psi^{\mu}_L\pa_z \psi^{\nu}_L +
\Phi_{\mu\nu\rho\sigma}\psi^{\mu}_L \psi^{\nu}_L \psi^{\rho}_L \psi^{\sigma}_L
\label{Xlv}
\comma
\eeq
with a similar formula for $\tilde{X}_R$.
The $\psi$'s in (\ref{Xlv}) are the left handed fermions in the sigma-model.
This $\tilde{X}$ and its superpartner $\tilde{M}$
together with $T$ and $G$ make a closed algebra,
and we will refer to it as the Ising superconformal algebra
(ISCA).

Let us impose now the boundary conditions.
In order to preserve the $N=1$ SCA we require
\beq
T_L=T_R,~~~~~G_L=\pm G_R
\stop
\label{N1}
\eeq
Also, we have to preserve a linear combination of the left and right
spectral flow operators.
The ISCA algebra implies
that
\beq
\tilde{X}_L = \tilde{X}_R,~~~~~\tilde{M}_L=\pm \tilde{M}_R
\stop
\label{XM}
\eeq
Thus, there is only one type of boundary condition in this case.

The conditions (\ref{N1}) are solved in the large volume limit by
\beq
\pa X^{\mu} = R^{\mu}_{\nu}\bar{\pa}X^{\nu},~~~~~
\psi_L^{\mu} = \pm R^{\mu}_{\nu}\psi_R^{\nu}
\comma
\label{geom1}
\eeq
where
\beq
g_{\mu\nu}R^{\mu}_{\rho}R^{\nu}_{\sigma} = g_{\rho\sigma}
\label{geom2}
\stop
\eeq
Here $X^{\mu}$ and $\psi^{\mu}$ denote coordinates and vielbein one-forms
on the manifold.
The eigen-vectors of $R$ with eigen-values $(-1)$ give the Dirichlet boundary
condition and thus correspond to the directions normal to the D-brane.
As noted above,
in the large volume limit $\tilde{X}$ takes the form (\ref{Xlv}).
Using (\ref{geom1}),(\ref{geom2}) and (\ref{Xlv}) we see that
the condition  (\ref{XM}) reads
\beq
\Phi_{\mu\nu\alpha\beta}R^{\mu}_{\rho}R^{\nu}_{\sigma}
R^{\alpha}_{\gamma}R^{\beta}_{\delta}=\Phi_{\rho\sigma\gamma\delta}
\stop
\label{cay}
\eeq
Remembering that $\Phi$ is self-dual we see that the geometrical
content of (\ref{cay}) is that $\Phi$ is the volume form of the supersymmetric
cycle.
Thus it is a Cayley submanifold as expected.

Since the boundary condition corresponding to the Cayley submanifold
preserves a linear combination of the spectral flow operators we see that
the $(2,0)$ space-time supersymmetry of type IIB compactified
on $Spin(7)$ holonomy manifold is broken by a D-brane wrapping on a Cayley
submanifold to $(1,0)$.

\section{$SU(4)$ holonomy}

\subsection{SCFT framework}

A Calabi-Yau  4-fold
with $SU(4)$ holonomy posses two covariantly constant spinors of the same
chirality.
Thus, there exist two corresponding
spectral flow
operators $\Psi_L$ and $\Psi_L^*$ of dimension $\frac{1}{2}$.
Combined with
$\Psi_R$ and
$\Psi_R^*$ we have four
spectral flow operators which means that type IIB string
compactified on a Calabi-Yau 4-fold to $1+1$ dimensions has $(4,0)$
space-time supersymmetry.

As we noted, supersymmetric cycles of special Lagrangian and holomorphic
types
are
associated with $A$ and $B$ types of boundary conditions respectively
\cite{OOY}.
These boundary
conditions preserve two linear combinations of
the spectral flow operators
$\{\Psi_L, \Psi_L^*, \Psi_R, \Psi_R^*\}$ which implies that
wrapping D-branes on these cycles breaks half of the space-time supersymmetry.
Thus, the $(4,0)$ space-time supersymmetry is broken down to $(2,0)$.

One can also define for Calabi-Yau 4-folds
an $S^1$ family of Cayley calibrations by
\beq
\Phi_{\theta} = \frac{1}{2}k^2 + Re(e^{i\theta}\Omega)
\stop
\label{phit}
\eeq
Since $k^2$ vanishes on special Lagrangian submanifolds, and
$Re(e^{i\theta}\Omega)$ vanishes on complex submanifolds,
 the calibration (\ref{phit}) includes the special Lagrangian
and complex calibrations as special cases. However a general Cayley
submanifold is neither special Lagrangian nor complex.
Note also that it cannot be simultaneously special Lagrangian and complex,
since
the K\"ahler form vanishes on Lagrangian submanifolds. Indeed we expect a
special Lagrangian cycle and a complex 4-cycle to intersect transversely
(at points) in
the 4-fold.

In this section we study the Cayley type supersymmetric cycle.
We will show that
the boundary condition associated with the Cayley submanifold
preserves only one linear combination
of the four spectral flow operators and thus only a quarter of the
space-time  supersymmetry.

In view of the previous section, we know that we have to preserve the
spin 2 operator $\tilde{X}$ corresponding to the energy momentum tensor
of the preserved  spectral flow operator.
In order
to formulate the boundary condition we embed the ISCA algebra in the
$N=2$ SCA as
\beq
   T = T_{N=2},~~~~~~
   G = G_{N=2}^+  +  G_{N=2}^-,~~~~~~
\tilde{X} = \frac{1}{2}J^2 + Re(e^{i\theta}\Omega)
\comma
\label{alg}
\eeq
with
   $\tilde{M}$  as the superpartner of $\tilde{X}$.
In the large volume limit $\tilde{X}$ takes the form
\beq
\tilde{X}_L = \frac{1}{2}g_{\mu\nu} \psi^{\mu}_L\pa_z \psi^{\nu}_L +
(\frac{1}{2}k^2 + Re(e^{i\theta}\Omega))
_{\mu\nu\rho\sigma}\psi^{\mu}_L \psi^{\nu}_L \psi^{\rho}_L \psi^{\sigma}_L
\label{Xcy}
\comma
\eeq
where we used the large volume
limit expressions
$J_L=g_{\mu\nu}\psi^{\mu}_L\psi^{\nu}_L$ and $\Omega =
\Omega_{\mu\nu\rho\sigma}\psi^{\mu}_L \psi^{\nu}_L \psi^{\rho}_L
\psi^{\sigma}_L$.
Equation (\ref{Xcy}) is expected since as noted in
(\ref{Xlv}), $\tilde{X}$ consists of two parts: The energy momentum tensor for
the fermions and the Cayley calibration form, and the latter
is given in (\ref{phit}).
Note that in fact (\ref{alg}) defines an $S^1$ family of embeddings
as suggested by (\ref{phit}).

Let us also verify that  $\tilde{X}$  is indeed the energy-momentum
tensor for the Ising model.
One way to see that
is to bosonize the $U(1)$ current $J = i \pa_z \phi$ and use
$\Omega = e^{i\phi}$.
Thus,
\beq
   \tilde{X}_L = \frac{1}{2}(\pa_z \phi)^2 + {\rm cos}(\phi+\theta)
\label{TPCO}
\stop
\eeq
Combining the two spectral flow operators
as  $e^{i (\phi+\theta)} = \Psi_1 + i \Psi_2$ we see that
$\tilde{X}_L = \Psi_1 \pa_z \Psi_1$,
namely $\tilde{X}_L$ is the energy-momentum tensor for the
Majorana-Weyl spinor $\Psi_1
$ with $c={1 \over 2}$\footnote{That the energy-momentum tensor of the Ising
model is given by (\ref{TPCO}) was shown in \cite{eguhi}.}.

The boundary condition that corresponds
to a Cayley submanifold which is neither special Lagrangian nor
complex is that of (\ref{N1}) and (\ref{XM}).
Thus, as we discussed, we are only preserving the
energy-momentum tensor for one linear combination of spectral flow operators
 and break the
rest of the $N=2$ SCA. This leaves us with one quarter of the supersymmetry.
The $S^1$ family of Cayley calibrations corresponds to the choice of the
preserved  linear
combination of the spectral flow operators.

Until now the only known way for D-branes to break more than half of the
space-time supersymmetry was to use a configuration of intersecting branes
\cite{intersect}. The Cayley
submanifold provides the first and the only example of a supersymmetric
cycle on which a single wrapped D-brane breaks three quarters of the
space-time supersymmetry.

\subsection{Low energy effective action framework, I}

In this and the following subsections we will use the low effective action
framework
in order to derive the conditions for
supersymmetric cycles in Calabi-Yau 4-folds. This will make the space-time
interpretation of the previous results manifest.
To derive the conditions for having a
supersymmetric 4-cycle, we consider the 3-brane
of the ten-dimensional type IIB theory which wraps a 4-cycle of the
Calabi-Yau 4-fold.
The 3-brane solution of the type IIB theory was discovered in
\cite{hostr} and its static gauge field content is described by
an abelian $D=4$, $N=4$ vector multiplet
\cite{threeb}. However, the covariant
3-brane action with the local $\kappa$ symmetry
has not been constructed so far, so that it is hard
to make a rigorous analysis along the lines of \cite{BBS}.

Alternatively, one may take the point of view that the low energy
effective action for the Euclidean D3-brane is the ``twisted''
$N=4$ Yang-Mills theory \cite{vw} and count the number of unbroken
supersymmetries by studying how the twisting is realized
on the Cayley submanifold. According to \cite{bsv}, the twisting
structure can be understood from the behavior of the normal bundle
of the submanifold.  For special Lagrangian
submanifolds, the $SU(4)$ global symmetry of $N=4$ decomposes
as $(2,1)\oplus (1,2)$ under the Lorentz group $SU(2)\times SU(2)$,
which leads to 2 unbroken supersymmetries\footnote{
The case of complex submanifolds is a bit different,
and does not fall into the classification given in \cite{vw} since the
normal bundle is not trivial before twisting.  Nevertheless, in this
case
too there are 2 covariantly constant spinors.}.

For a Cayley submanifold in a manifold with
$Spin(7)$-holonomy, the normal bundle can be written in the form
\cite{McLean} $\SS_-\otimes {\cal F}$, where ${\cal F}$ is a certain rank two
vector bundle
on the Cayley submanifold.  As pointed out in \cite{bsv}, if the vector
bundle ${\cal F}$ is trivial then the twisting is the one for which the global
$SU(4)$ symmetry
decomposes as $(1,2)\oplus (1,1)\oplus (1,1)$ under the Lorentz group;
this leaves 1 unbroken supersymmetry.  It remains to verify that in our
situation---a Cayley submanifold of a manifold with $SU(4)$ holonomy which
is neither a complex nor a special Lagrangian submanifold---the
bundle ${\cal F}$ is trivial.

The structure of the normal bundle of a Cayley submanifold in the $Spin(7)$
holonomy case is analyzed in some detail by McLean \cite{McLean}.
The half-spin representations of $Spin(8)$ are eight-dimensional; if we
fix a spinor $\sigma$ in one of the representation spaces, its stabilizer
is isomorphic to $Spin(7)$.  Projecting that copy of $Spin(7)$ to the
vector representation of $Spin(8)$ produces the holonomy representation
$Spin(7)\to SO(8)$.
If the actual holonomy is $SU(4)\cong Spin(6)$, there will be an embedding
of $Spin(6)$ in $Spin(7)$, determined by a second spinor
$\sigma'$ of which $Spin(6)$ is the stabilizer (within $Spin(7)$).

Given a Cayley 4-plane $\xi$, there are quaternionic structures on
the 4-planes $\xi$ and $\xi^\perp$
such that the stabilizer $G_\xi$ of $\xi$ in
$Spin(7)$  can be written as
\beq
G_\xi=(Sp(1)_L\times Sp(1)^\perp_L\times Sp(1)^{diag}_R)/\{\pm(1,1,1)\} ,
\eeq
where $Sp(1)_L$ and $Sp(1)_R$ are the two natural subgroups of $SO(\xi)$
given by the left and right actions of the unit quaternions,
$Sp(1)^\perp_L$ and $Sp(1)^\perp_R$ are the corresponding subgroups of
$SO(\xi^\perp)$, and $Sp(1)^{diag}_R$ is the diagonal subgroup of
$Sp(1)_R\times Sp(1)^\perp_R$.

In terms of the embedding $Spin(7)\subset Spin(8)$, the group
$Sp(1)_L\times Sp(1)_L^\perp$
is the stabilizer of a 4-plane $\eta$ of spinors orthogonal to $\sigma$, and
$Sp(1)_R^{diag}$ is the intersection of the stabilizer of $\eta^\perp$ with
$Spin(7)$.  If we choose an embedding $Spin(6)\subset Spin(7)$ corresponding
to a spinor $\sigma'$, then there are three possibilities for the
intersection of $Spin(6)$ with $G_\xi$:
\begin{enumerate}
\item
$\sigma'\in \eta^\perp$, in which case
$Spin(6)\cap G_\xi=Sp(1)_L\times Sp(1)_L^\perp\times U(1)$ with
$U(1)\subset Sp(1)_R^{diag}$, and

\item
$\sigma'\in \eta$, in which case 
$Spin(6)\cap G_\xi=SU(2)\times Sp(1)_R^{diag}$
with $SU(2)\subset Sp(1)_L\times Sp(1)_L^\perp$ conjugate to the diagonal
embedding

\item
$\sigma'$ generic, in which case $Spin(6)\cap G_\xi=SU(2)\times U(1)$ with
$SU(2)\subset Sp(1)_L\times Sp(1)_L^\perp$ conjugate to the diagonal embedding
and $U(1)\subset Sp(1)_R^{diag}$.
\end{enumerate}
In case 1, the orbit of $\xi$ under $Spin(6)=SU(4)$ takes the form
 $Spin(6)/(Spin(6)\cap G_\xi)\cong SU(4)/S(U(2)\times U(2))$, from which
it is clear that $\xi$ is a complex subspace of $\RR^8=\CC^4$.
In case 2, the orbit of $\xi$ takes the form
 $Spin(6)/(Spin(6)\cap G_\xi)\cong SU(4)/SO(4)$ which implies that
$\xi$ is a special Lagrangian 4-plane.  Finally, in case 3 the orbit of
$\xi$ takes the form $Spin(6)/(Spin(6)\cap G_\xi)\cong SU(4)/(SU(2)\times
U(1))$
and has dimension $11$ (different from the previous cases), so $\xi$ must
be a Cayley 4-plane which is neither a complex  nor special Lagrangian
subspace.

    To make contact with the SCFT approach in section 3.1, 
we can count the number of supersymmetries preserved, or 
equivalently, the number of supersymmetries broken.  Each 
of the latter would generate a goldstino, i.e. a fermion zero modes in 
in the low energy effective super Yang-Mills action.  
In the present cases, they correspond to covariantly constant 
spinors.  Generically, there are no more covariantly constant 
spinors.  Thus the number of unbroken spacetime supersymmetries 
is equal to that of covariantly constant spinors in the fermion 
bundles of the low energy action.  
Using the intersection of $Spin(6)$ with $G_\xi$ given above, we 
find them to be 2, 2, and 3 for complex, 
special lagrangian, and Cayley submanifolds respectively.  
Since the total number of spacetime supersymmetries in 
SU(4) compactification of type II string theory is 4, 
this reproduces the counting given in section 3.2.

Following \cite{McLean}, when the holonomy is $Spin(7)$,
the vector bundle ${\cal F}$ in the normal bundles to a 
Cayley submanifold is the rank two bundle naturally associated
to the principal bundle $Sp(1)_L^\perp$.
(In McLean's $p$, $q$, $r$ notation, $p\in Sp(1)_L$, $q\in Sp(1)_R^{diag}$
and $r\in Sp(1)_L^\perp$.)  Nonetheless, one can show, along the 
same line of reasoning employed above, that 
it is supersymmetric even when ${\bar F}$ is nontrivial.

\subsection{Low energy effective action framework, II}

Although a covariant 3-brane action with the local $\kappa$-symmetry
is not yet  known, it is not difficult to guess what its symmetry
structure should be if there is one. By making a reasonable assumption
on the symmetry structure
of the would-be covariant action, one may formally extend the
analysis of \cite{BBS} to the present case, and give another derivation
of the results of the previous subsection.
The conditions satisfied by a supersymmetric 4-cycle
are expressed in terms of a holomorphic $(4,0)$-form:
\beq
\Omega=\frac{1}{4!} \Omega_{abcd} (X) dX^a \wedge dX^b \wedge dX^c
\wedge dX^d \comma
\eeq
and the K\"ahler form
\beq
k_{a {\bar b}}=i g_{a {\bar b} }
\stop
\eeq
Here $a,{\bar b}$ are holomorphic and anti-holomorphic indices and
$X^i$ are coordinates on the Calabi-Yau.

As discussed in  subsection 3.1, an
eight-manifold with $SU(4)$ holonomy has two covariantly constant
eight-dimensional Majorana-Weyl spinors $\epsilon_1$ and $\epsilon_2$,
with the same chirality. Changing the
chirality of both spinors would correspond to reversing the orientation.
We can combine these real spinors into a complex
spinor $\epsilon_+=\epsilon_1 +i \epsilon_2$,
whose normalization can be chosen
as $\epsilon_+^{\dagger}\epsilon_+=1$. The K\"ahler form
is expressed in terms of this spinor as:
\beq
k_a^{\,\,\, b} =i \epsilon_+^{\dagger} \gamma_a^{\,\,\, b} \epsilon_+
\stop
\label{aiii}
\eeq
In general,
$\gamma_{m_1 \dots m_n}$ is the completely antisymmetrized product of $n$
eight-dimensional gamma matrices containing a factor $1/n!$.
{}From (\ref{aiii}) it can be easily seen that $\gamma_{a}$
acts as an annihilation operator
\beq
\gamma_{a} \epsilon_+=\gamma_{\bar a} \epsilon_-=0
\comma
\eeq
where $\epsilon_-=(\epsilon_+)^*$.
The holomorphic 4-form relates $\epsilon_+$ and $\epsilon_-$:
\beq
\gamma_{abcd} \epsilon_-=\Omega_{abcd} \epsilon_+
\stop
\eeq

Using standard properties
of gamma matrices,
it can be shown that the following formulas hold
\beqar
\gamma_{ a {\bar b}{\bar c}{\bar d} }\epsilon_+ &= &- 3i k_{ a [{\bar b}}
\gamma_{{\bar c}{\bar d}] }\epsilon_+ \comma
\nonumber\\
\gamma_{{\bar a} {\bar b} cd} \epsilon_+ &=&
3k_{[ {\bar a} c }k_{{\bar b} d]}\epsilon_+
\stop
\eeqar
Similar equations involving the spinor $\epsilon_-$ can be obtained
after complex conjugation.

Although the covariant action for the 3-brane is yet to be constructed,
it should be natural to assume, by extending the analysis in \cite{BBS},
 that
the 3-brane would preserve the supersymmetries generated by
ten-dimensional spinors
$\epsilon$ if they solve\footnote{
By comparing with the analysis of section 3.2 using the $N=4$
Yang-Mills theory, we note that, in the case of a special Lagrangian
submanifold, we have to take into account both projection operators
$P_+$ and $P_-$.} 
\beq
P_- \epsilon =\frac{1}{2}\left( 1-\frac{1}{4!} \epsilon^{\m\n\r\s}
\pa_{\m} X^M \pa_{\n} X^N\pa_{\r}X^P \pa_{\s} X^Q \G_{MNPQ} \right)
\epsilon=0
\stop
\label{avii}
\eeq
In the covariant formulation, this would be a condition for
the local $\kappa$-transformation to compensate for
the global supersymmetry generated by $\epsilon$.
Here $M,N=1,\dots,10$ are ten-dimensional indices, $\m,\n=1,\dots,4$
are the
worldbrane indices \footnote{This notation is different than the one used
in the previous sections where the distinction between worldbrane indices and
Calabi-Yau indices was encoded in the eigen-vectors of the $R$ matrix.},
$\G_M$
are the ten-dimensional gamma matrices and $X^M$ is the bosonic
part of the 3-brane configuration. In the above formula
we have introduced the projection operator $P_{-}$ which
is hermitian and satisfies $P_{-}^2=P_{-}$.

Let us introduce the eight-dimensional spinor:
\beq
\epsilon_{\theta} =e^{-i \theta/2} \epsilon_+ +e^{i \theta/2} \epsilon_-
\stop
\eeq
The spinor $\epsilon$ can then be written in the form $\epsilon =\lambda
\epsilon_{\theta}$, where $\lambda$ is a two-dimensional Majorana
spinor. The simplest way to find solutions of the equation $P_- \epsilon
_{\theta}=0$, is to derive a Bogomolnyi bound (which will be closely
related to the calibration condition, as we have remarked earlier).
This bound is saturated
if and only if the 4-cycle is supersymmetric. This
implies that the 3-brane has minimized its volume. The bound can be
derived from the inequality
\beq
\int d^4 \s \sqrt{h} (P_- \epsilon_{\theta} )^{\dagger}
P_- \epsilon_{\theta} \geq 0
\stop
\eeq
In the above formula $P_-$ is constructed from eight-dimensional
gamma matrices and $h$ is the induced metric on the 3-brane.
After a straightforward computation we
obtain the result:
\beq
V_4 \geq \frac{1}{2} \int k \wedge k +{\rm Re} (
e^{i \theta} \int \Omega )
\comma
\label{aix}
\eeq

Comparing to the Cayley calibration (\ref{phit}) we see that
the manifolds which saturate this bound
correspond to the $S^1$ family
of Cayley submanifolds, that we previously found.
As we already pointed out, Lagrangian submanifolds and complex submanifolds
are special cases of Cayley geometries. The complex submanifolds found herein
coincide precisely with the expression found in \cite {BBS} for
the supersymmetric 3-brane wrapping a 4-cycle of a Calabi-Yau 3-fold.

\subsection{Examples}

The simplest examples of supersymmetric
4-cycles can be found in flat
space\footnote{Here we will solve (\ref{aix}) pointwise. }.
Here:
\beqar
\Omega &=& dX^1 \wedge dX^2 \wedge dX^3 \wedge dX^4 \comma \nonumber\\
k &=& dX^1 \wedge dX^{\bar 1} +\dots  +dX^4 \wedge dX^{\bar 4}
\stop
\eeqar

An example of a Lagrangian submanifold
is the surface described by $X^i =X^{\bar i}$ for $i=1,\dots,4$.
In that case equation (\ref{aix}) is saturated because the pullback of
$\Omega$ satisfies
\beq
\pa_{\a} X^m \pa_{\b} X^n \pa_{\g} X^p \pa_{\delta} X^q
\Omega_{mnpq} =\epsilon_{\alpha\beta\gamma\delta}
\comma
\eeq
while the pullback of $k$ vanishes.

A more complicated example, that is not in flat space, can be
found as a 4-cycle in the sextic hypersurface
\beq
\sum_{i=1}^6 (X^i)^6=0
\stop
\eeq
in $CP^5$. This 4-cycle is the four-dimensional
submanifold on which all the $X^{i}$'s are real \cite{Bryant,BBS}.

An example of a complex submanifold is
given by the surface described by
$X_3=X_4=0$. Here the pullback of $\Omega$ vanishes and the
pullback of $k\wedge k$ is
\beq
\pa_{\a} X^m \pa_{\b} X^n \pa_{\g} X^{\bar p}
\pa_{\delta} X^{\bar q} k_{m{\bar p}} k_{n{\bar q}}=2 \epsilon_{\a\b\g\delta}
\comma
\eeq
so that (\ref{aix}) is saturated.

An example of a Cayley geometry, for which both the
pullback of the holomorphic 4-form and the pullback of $k\wedge k $ are
non-vanishing is described by
$ X^2=\sqrt{2}e^{i \varphi} (X^1 +X^{\bar 1})$ and
$X^4=\sqrt{2}e^{i \varphi}(X^3+X^{\bar 3})$, for every value of the
angle $\varphi$. More generally, every Cayley plane that is neither special
Lagrangian nor holomorphic will give an example of this type.

\section{$G_2$ holonomy}

\subsection{SCFT framework }


Let $M$ be an seven-manifold. A $G_2$ structure on $M$ is given by
a closed $G_2$ invariant 3-form $\Phi$.
This defines a metric $g$ with
holonomy group $Hol(g) \subset G_2$. Such a metric is Ricci-flat.
Compact $G_2$ holonomy manifolds have been constructed in 
\cite{Joyce1,Joyce2}
in analogy with the $Spin(7)$ holonomy case
by resolving the singularities of $T^7/\Gamma$ orbifolds. Here $T^7$ is
equipped with a flat $G_2$ structure and $\Gamma$ is a finite group
of  isometries of $T^7$ preserving that structure.
On a $G_2$ holonomy manifold there exists one covariantly constant
spinor.
The 3-form $\Phi$ and its Hodge dual 4-form $^*\Phi$  define the associative
and coassociative calibrations respectively.


The extended symmetry algebra of sigma models on $G_2$ manifolds
has been constructed in \cite{sv}.
In addition to the stress tensor $T$ and its superpartner $G$,
it contains the superpartners $(K,\Phi)$ with spins $(2,\frac{3}{2})$
and $(X,M)$ with spins $(2,\frac{5}{2})$.
In the large volume limit, $\Phi$ corresponds to the associative
calibration 3-form and $X$ is the sum of
the coassociative
calibration 4-form  $^*\Phi$ and the stress tensors for
seven Majorana-Weyl fermions.
In analogy with the $Spin(7)$ holonomy case where we viewed
$\tilde{X}$ as the stress tensor corresponding to the dimension
$\frac{1}{2}$ spectral flow operator, here we can view $X$ as
the stress tensor corresponding to
the dimension $\frac{7}{16}$ spectral flow operator which is the spin
field of the $c=\frac{7}{10}$ tri-critical Ising model.

In addition to the $N=1$ boundary condition (\ref{N1}), the $G_2$ algebra
implies the boundary conditions
\beqar
\Phi_L &=& \Phi_R,~~~~~K_L=\pm K_R\comma
\nonumber\\
X_L &=& X_R,~~~~~M_L=\pm M_R
\stop
\label{g2bc}
\eeqar
In the large volume limit we have
\beq
\Phi_L = \Phi_{ijk}\psi^i_L \psi^j_L \psi^k_L,~~~~~~
X_L = \frac{1}{2}g_{ij} \psi^i_L\pa_z \psi^j_L +
^*\Phi_{ijkl}\psi^i_L \psi^j_L \psi^k_L \psi^l_L
\label{Xp}
\stop
\eeq
Thus the boundary conditions (\ref{g2bc}) take the form
\beq
\Phi_{ijk}R^i_lR^j_m
R^k_n=\Phi_{lmn},~~~~~~
^*\Phi_{ijkl}R^i_mR^j_n
R^k_oR^l_p=^*\Phi_{mnop}
\comma
\eeq
which geometrically mean that for the 3-cycle $\Phi$ is the volume form
while for a 4-cycle $^*\Phi$ is the volume form.
These are the associative and coassociative calibrated submanifolds.
Since the boundary conditions (\ref{g2bc}) impose one linear constraint
on the stress tensor operator corresponding to the spectral flow we see that
a brane wrapping on an associative or coassociative cycle preserves half of
the space time supersymmetry. Thus, the $(2,0)$ space-time supersymmetry
of a type IIB string compactified on a $G_2$ holonomy seven-manifold is
broken to $(1,0)$ by the brane.

\subsection{Low energy effective action framework}
Supersymmetric 3-cycles are defined as configurations for which
we can find a seven-dimensional spinor that satisfies
\beq
P_-\epsilon =\frac{1}{2} \left(1-\frac{i}{3!} \epsilon^{\mu\nu\rho}
\pa_{\mu}X^m \pa_{\nu}X^n \pa_{\rho} X^p \Gamma_{mnp} \right) \epsilon =0
\stop
\eeq
This expression is evaluated using the 3-form $\Phi$,
which appears in the expression
\beq
\gamma_{mnp} \epsilon =\Phi_{mnp} \epsilon
\stop
\eeq
{}From here we see that the only configuration that preserves
supersymmetry satisfies that the pullback of the 3-form is
proportional to the volume element:
\beq
\pa_{[\mu}X^m \pa_{\nu} X^n \pa_{\rho]} X^p \Phi_{mnp}
=\epsilon_{\mu\nu\rho}
\stop
\eeq
These are precisely the associative calibrations previously discussed.
On a manifold
with $G_2$ holonomy we can also have 4-cycles in which 
$^*\Phi$ pulls back to
the volume element. These cycles satisfy
\beq
P_-\epsilon =\frac{1}{2} \left( 1-\frac{1}{4!}
\epsilon^{\mu\nu\rho\sigma} \pa_{\mu}X^m \pa_{\nu} X^n \pa_{\rho}X^p
\pa_{\sigma} X^q \G_{mnpq} \right) \epsilon =0
\comma
\eeq
which is solved by the configuration
\beq
\pa_{[\mu} X^m \pa_{\nu} X^n \pa_{\rho} X^p \pa_{\sigma]} X^{q} 
{^*\Phi_{mnpq}}
=\epsilon_{\mu\nu\rho\sigma}
\stop
\eeq
Thus $^*\Phi$ corresponds to the coassociative calibration.
Both configurations break half of the supersymmetry.

\section{Mirror symmetry}
In the case of the 4-fold with $SU(4)$ holonomy, we may consider
the effect of mirror symmetry which exchanges $G_{N=2}^+$ and $G_{N=2}^-$
for the right mover.   As was shown in \cite{OOY}, mirror symmetry
exchanges the $A$ and $B$ types of boundary conditions.  Geometrically,
mirror symmetry
is realized on {\it pairs}\/ of Calabi--Yau manifolds which
define the same theory (but with opposite geometric identifications
of $G_{N=2}^{\pm}$).  Thus,
if $X$ and $Y$ are a pair of mirror manifolds, the special Lagrangian
submanifolds of $X$ are mapped to the complex submanifolds of $Y$,
and the complex submanifolds of $X$ are mapped to the special Lagrangian
submanifolds of $Y$.

Mirror symmetry for 4-folds has several new features which distinguish
it from the three-dimensional case \cite{GMP}.  Mirror symmetry is
 expected to map $H^4(X)=\bigoplus_p H^{p,4-p}(X)$ to 
$\bigoplus_p H^{p,p}(Y)$
and $\bigoplus_p H^{p,p}(X)$ to $H^4(Y)=\bigoplus_p H^{p,4-p}(Y)$; 
one of the new
features is that $H^{2,2}(X)$ appears in both of these spaces.  (These
spaces were referred to as the ``horizontal'' and ``vertical'' cohomology
in \cite{GMP}.)

 The special Lagrangian submanifolds of $X$ define classes in $H^4(X)$ which
lie
in the so-called {\it primitive cohomology}, that is, they are classes
which are orthogonal
to the K\"ahler class.  Since the classes of special Lagrangian
submanifolds are also classes in integer cohomology, the natural space
 to consider for these manifolds is
$H^4(X)_{prim}\cap H^4(X,\ZZ)$.  It is not clear how much of this space
will actually be represented by special Lagrangian submanifolds.

On the other hand, the complex submanifolds of $X$ define classes
which have Hodge type $(p,p)$ and are also integer cohomology classes;
the natural space to consider for them is
$\bigoplus_p H^{p,p}(X)\cap H^{even}(X,\ZZ)$.  The celebrated ``Hodge
conjecture''
in mathematics asserts that if we pass to $\QQ$-coefficients instead of
$\ZZ$-coefficients, then all classes in this space are represented by
complex submanifolds; it is not known if this conjecture 
holds for Calabi--Yau 4-folds.

We are thus faced with the situation of having an unknown subspace of
$H^4(X)_{prim}\cap H^4(X,\ZZ)$ represented by special Lagrangian submanifolds,
and an unknown subspace of $\bigoplus_p H^{p,p}(X)\cap H^{even}(X,\ZZ)$
represented
by complex submanifolds.  In fact, it is quite possible that the appropriate
pieces of these subspaces fall short of filling out all of $H^{2,2}(X)$ (even
though both will contribute subspaces of $H^{2,2}(X)$).  Cayley submanifolds
provide another potential source of cohomology classes which could help to
fill out $H^{2,2}(X)$: it may be that some of the classes which cannot be
represented by either special Lagrangian or complex submanifolds will instead
be represented by Cayley submanifolds.

Such a possibility meshes well with
mirror symmetry: we observe that the mirror of a Cayley submanifold will
be another Cayley submanifold.  (This is because any D-brane on $X$---which
defines
some type of boundary condition for open strings---should map to
a D-brane on $Y$.)  If the first Cayley submanifold is neither special
Lagrangian nor a complex submanifold, then since it preserves only 1/4
of the supersymmetry, its mirror will have the same
property.  It would be interesting to find explicit examples of this
phenomenon.

Finally, we would like to mention an implication
for mirror symmetry
in higher dimensions that becomes evident by considering the spectrum
of BPS soliton states.
Recently Strominger, Yau and Zaslow \cite {syz} showed that every Calabi-Yau
3-fold that has a mirror admits a supersymmetric $T^3$-fibration.
The basic assumption of this argument is quantum mirror symmetry
\cite {andy,am,BBS,morrison},
where the isomorphism between the type IIA theory compactified on a
3-fold $X$ and the IIB theory compactified on the mirror $Y$ of $X$
is extended to the non-perturbative BPS states in $D=4$.
Since these BPS states are constructed as D-branes, the
quantum mirror symmetry is actually a consequence of the
classical mirror symmetry of the bulk CFT \cite{OOY}.
It is then natural to wonder if the previous argument can be extended to
higher dimensional Calabi-Yau manifolds. Some precise mathematical
aspects of this generalization have been recently considered in \cite {mo}.
We consider the type IIA theory compactified on a large Calabi-Yau $n$-fold
$X$ and its mirror $Y$. Quantum mirror symmetry implies that both theories
are isomorphic. On the $X$ side there are `BPS states' in $D=(10-2n)$
\footnote{Rigorously, the notion of a
BPS state that carries electro-magnetic charge in $D \leq 2$ is not
well defined due to the occurrence of infrared divergences.
(We thank A.~Strominger for pointing this out to us).
The problem can be better formulated by translating
the result of \cite{syz} into the CFT language.}
which arise
from the ten-dimensional 0-brane. These states arise from a supersymmetric
$n$-brane wrapping a $n$-cycle in $Y$. This $n$-cycle corresponds to a
special Lagrangian submanifold. This is because the 0-brane
corresponds to B-type boundary conditions and by mirror symmetry these
are transformed to the A-type boundary conditions that correspond
to the special Lagrangian submanifold \cite {OOY}.
Extending the arguments of \cite {syz} to $n$-folds, we arrive at the
conclusion that the $n$-cycles corresponding 
to special Lagrangian submanifolds
are toroidal. This leads us to the
conclusion that every Calabi-Yau $n$-fold that has a mirror admits a
supersymmetric $T^n$-fibration. This suggests that the mirror
symmetry for the $n$-fold is
equivalent to a T-duality on the $T^n$-fibers.

\section*{Acknowledgements}
We would like to thank
R.~Bryant, G.~Moore, R.~Plesser, J.~Polchinski,
A.~Schwimmer,  A.~Strominger, C.~Vafa,
and Y.~Vlassopoulos for useful discussions.
D.R.M. and H.O. are grateful for the hospitality of the Aspen Center for
Physics
during the final stages of this project.
The work of K.B. was supported by NSF grant PHY89-04035. The work of  M.B.
was supported by DOE grant DOE-91ER40618.
The work of D.R.M. was  supported in part by the National Science Foundation
under grant DMS-9401447.
The work of H.O. was supported in part by NSF grant PHY-951497
and DOE grant
DE-AC03-76SF00098. Y.O. is partially supported by the Israel 
Science Foundation
through
the Center for the Physics of Basic Interactions.
Z.Y. is supported by a Graduate Research Fellowship
of the U.S. Department of Education.

\end{document}